\documentclass[manuscript]{acmart}

\AtBeginDocument{%
  \providecommand\BibTeX{{%
    \normalfont B\kern-0.5em{\scshape i\kern-0.25em b}\kern-0.8em\TeX}}}

\makeatletter
\newcommand\footnoteref[1]{\protected@xdef\@thefnmark{\ref{#1}}\@footnotemark}
\makeatother

\newcommand\AICodeGenerator{AI code generator}
\newcommand\AICodeGenerators{AI code generators}
\newcommand\AICodeGeneration{AI code generation}
\newcommand\AICodeGenerationTools{AI code generators}

\usepackage{menukeys}
\newcommand\thetabkey{\keys{Tab}}

\usepackage{caption}
\usepackage{subcaption}
\usepackage{multirow}

\copyrightyear{2023}
\acmYear{2023}
\setcopyright{rightsretained}
\acmConference[TOCHI]{ACM Transactions on Computer-Human Interaction}
\acmBooktitle{ACM Transactions on Computer-Human Interaction}
\acmDOI{11.1111/1111111.1111111}
\acmISBN{111-1-111-1111-1/11/11}

\begin{document}

\title[Usability and Interactions with Copilot for Novice Programmers]{``It's Weird That it Knows What I Want'': Usability and Interactions with Copilot for Novice Programmers}

\author{James Prather}
\orcid{0000-0003-2807-6042}
\affiliation{
  \institution{Abilene Christian University}
  \city{Abilene, Texas}
  \country{USA}
}
\email{james.prather@acu.edu}

\author{Brent N. Reeves}
\email{brent.reeves@acu.edu}
\orcid{0000-0001-5781-1136}
\affiliation{%
  \institution{Abilene Christian University}
  \city{Abilene, Texas}
  \country{USA}
}

\author{Paul Denny}
\orcid{0000-0002-5150-9806}
\affiliation{
  \institution{The University of Auckland}
  \city{Auckland}
  \country{New Zealand}
}
\email{paul@cs.auckland.ac.nz}

\author{Brett A. Becker}
\orcid{0000-0003-1446-647X}
\affiliation{
  \institution{University College Dublin}
  \city{Dublin}
  \country{Ireland}
}
\email{brett.becker@ucd.ie}

\author{Juho Leinonen}
\orcid{0000-0001-6829-9449}
\affiliation{
  \institution{Aalto University}
  \city{Espoo}
  \country{Finland}
}
\email{juho.2.leinonen@aalto.fi}

\author{Andrew Luxton-Reilly}
\orcid{0000-0001-8269-2909}
\affiliation{
  \institution{The University of Auckland}
  \city{Auckland}
  \country{New Zealand}
}
\email{a.luxton-reilly@auckland.ac.nz}

\author{Garrett Powell}
\orcid{0000-0002-3221-7015}
\affiliation{%
  \institution{Abilene Christian University}
  \city{Abilene, Texas}
  \country{USA}}
\email{gbp18a@acu.edu}

\author{James Finnie-Ansley}
\orcid{https://orcid.org/0000-0002-4279-6284}
\affiliation{
  \institution{The University of Auckland}
  \city{Auckland}
  \country{New Zealand}
}
\email{james.finnie-ansley@auckland.ac.nz}

\author{Eddie Antonio Santos}
\orcid{}
\affiliation{
  \institution{University College Dublin}
  \city{Dublin}
  \country{Ireland}
}
\email{eddie.santos@ucdconnect.ie}

\renewcommand{\shortauthors}{Prather et al.}

\begin{abstract}

Recent developments in deep learning have resulted in code-generation models that produce source code from natural language and code-based prompts with high accuracy.  This is likely to have profound effects in the classroom, where novices learning to code can now use free tools to automatically suggest solutions to programming exercises and assignments. However, little is currently known about how novices interact with these tools in practice. We present the first study that observes students at the introductory level using one such code auto-generating tool, Github Copilot, on a typical introductory programming (CS1) assignment. Through observations and interviews we explore student perceptions of the benefits and pitfalls of this technology for learning, present new observed interaction patterns, and discuss cognitive and metacognitive difficulties faced by students. We consider design implications of these findings, specifically in terms of how tools like Copilot can better support and scaffold the novice programming experience.

\end{abstract}




\maketitle

%
%
\section{Introduction}
\label{sec:intro}
Introductory programming courses typically require students to write many programs~\cite{allen2021concise}. Teachers design programming exercises to facilitate and improve student learning; however, students are not always appropriately oriented to their learning, often focusing on completing tasks as quickly as possible. Therefore, in the context of these programming exercises, teachers and students often have competing user needs. These differing needs converge when students find themselves stuck and unable to complete the tasks required. There is plenty of evidence that students struggle to develop effective plans~\cite{soloway1986learning} and to implement plans that are developed~\cite{denny2012all, kazerouni2019student}. This is a frustrating experience for students~\cite{becker2018fix} that can limit their learning progress, and may result in undesirable behaviors such as copying~\cite{hellas2017plagiarism}. To maintain progress and positive learning experiences, an outcome desirable for both teachers and students, there is a need to support students who are stuck~\cite{McCartney2007unstuck}.  



Unfortunately, teachers are not always available to provide this support. Static learning resources such as tutorials, guides, textbooks, and tools such as IDEs do not provide contextualized interactive support. There have been numerous attempts to provide more contextualized help to students learning to program, although effective programming support remains a challenge~\cite{keuning2018systematic}. Intelligent tutoring systems~\cite{crow2018ITS} provide adaptive feedback to students depending on their performance, but such systems typically guide students through pre-constructed tasks and do not support student learning in more general (in-the-wild) environments. More recently, automated hint generation systems have been employed to generate hints for any programming exercises~\cite{McBroom2021HINTS}. Although such systems do not require the feedback on tasks to be manually designed, they do need to be deployed in environments that have access to historical student performance data to determine which approaches are more successful for a given exercise~\cite{Mahdaoui2022hints}, which limits their utility~\cite{keuning2018systematic}. None of the current automated approaches are able to provide contextual support for any programming problem on which a student is working.  Students (and their teachers) could benefit from automated support within a standard IDE that relates to the problem they working on. Large language models (LLMs) may fulfill this need.

In this paper, we explore how students used an LLM tool when engaged in a programming task, from a teaching and learning perspective. We study the use of GitHub Copilot -- an IDE plugin powered by the LLM Codex. Copilot is easily accessible to novices, is free for students, and operates as a plug-in directly in popular development environments. Existing work has not explored how students interact with LLM-based tools to support their progress in programming education. This paper extends knowledge in this field by presenting the first study that observes students at the first-year university level using Copilot on a typical assignment in an introductory programming course (often generically called ``CS1''~\cite{becker201950}). In particular, we were interested in capturing the novel experience of interacting with a new tool for the first time. We triangulate observations of novice programmers with interviews that explore their perceptions of the benefits and dangers of this technology for learning.  

We find that most students perceived that Copilot would help them write code faster, while also expressing concerns about not understanding the auto-generated code and becoming reliant on the tools -- concerns also held by educators~\cite{finnieansley2022robots, chen2021codex}. We observed two new interaction patterns. The first was when students guided Copilot by utilizing its auto-generated code prompts, \textit{shepherding} it toward a solution instead of focusing on writing code from scratch and integrating Copilot's suggestions. The second was when some students were moved along by some of Copilot's incorrect suggestions, \textit{drifting} from one to the next 
and therefore becoming lost. We also observed that students struggled with both cognitive and metacognitive difficulties when using the tool. Finally, we present ethical considerations and design guidelines based on these insights.

\begin{figure}
     \centering
     \begin{subfigure}[t]{0.49\textwidth}
     \centering
         \includegraphics[width=1\textwidth]{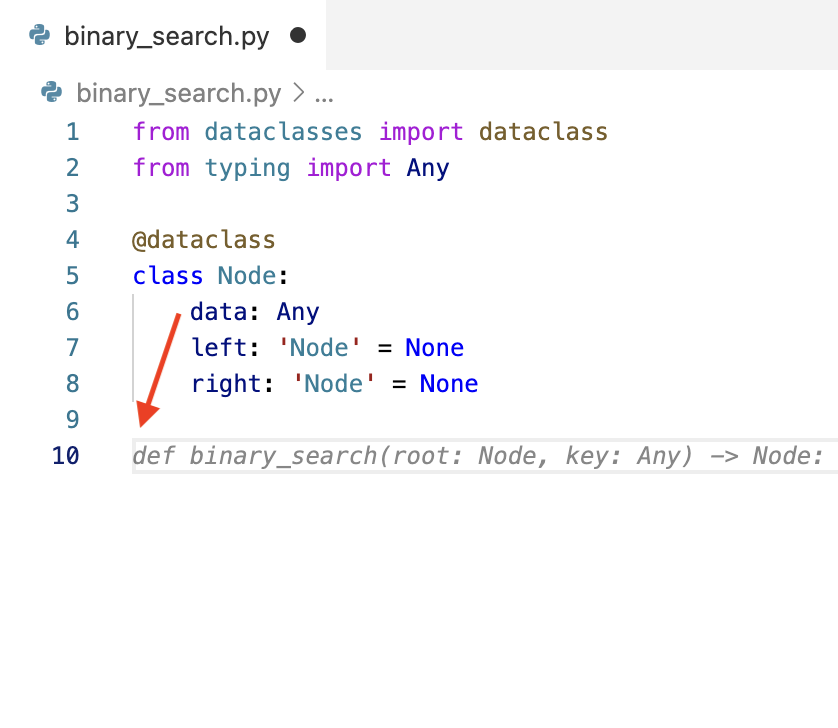}
         \caption{A programmer has entered code up to the cursor (red arrow) at the beginning of line 10 with minimal help from Copilot before this point. The gray italicized code is a Copilot suggestion.}
         \label{copilot-at-work-a}
     \end{subfigure}
     \hfill
     \begin{subfigure}[t]{0.49\textwidth}
         \centering
         \includegraphics[width=1\textwidth]{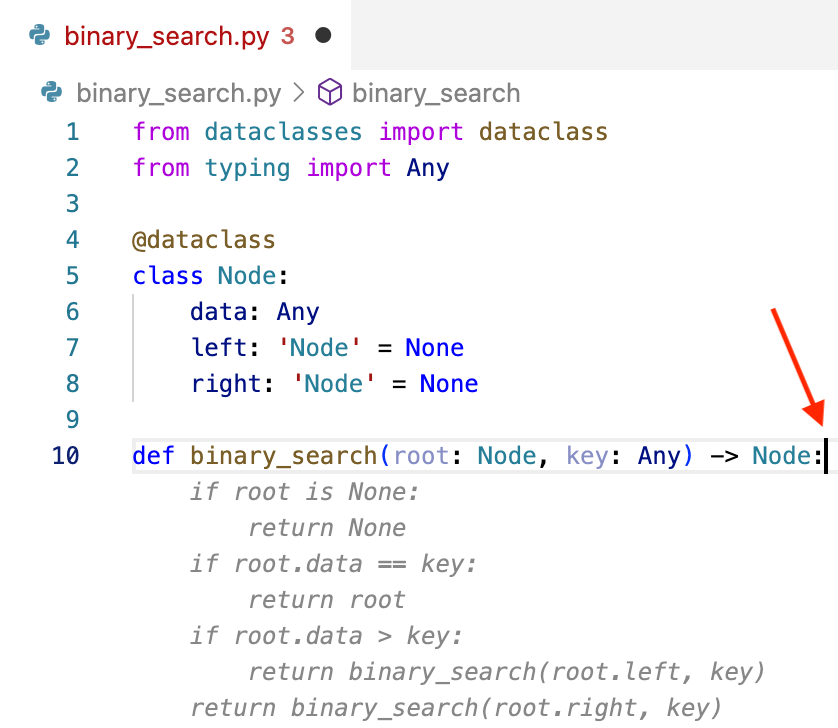}
         \caption{The programmer has accepted the suggestion in (a) by pressing tab, advancing the cursor to the end of line 10 (red arrow), just after the code that was suggested. The gray italicized code is the next Copilot suggestion
         .}
         \label{copilot-at-work-b}
     \end{subfigure}
     \caption{Copilot suggesting code to a programmer. Several videos are available at \href{https://github.com/features/copilot}{github.com/features/copilot}.}
        \label{copilot-at-work}
\end{figure}

\subsection{Background}
\label{sec:background}
In 2021 OpenAI released Codex~\cite{chen2021codex}, a version of GPT-3 trained on billions of lines of Python code from public GitHub repositories.\footnote{\label{codex}\href{https://openai.com/blog/openai-codex/}{openai.com/blog/openai-codex}}
Codex is conversant in both natural and programming languages. It is most proficient in English and Python but can also work in other natural languages such as Spanish and programming languages including Go, JavaScript, Perl, PHP, Ruby, Shell, Swift, and TypeScript.\footnoteref{codex}

GitHub Copilot uses Codex to suggest code in real-time based on code that has been entered by the user.\footnoteref{copilot} Copilot was moved out of technical preview in June 2022 and is now available for free to students as a plug-in for IDEs such as Visual Studio Code and JetBrains. Copilot is billed as ``Your AI pair-programmer''\footnote{\label{copilot}\href{https://github.com/features/copilot}{github.com/features/copilot}}---an intentional reference to pair programming, a  well-known programming education approach~\cite{mcdowell2002effects}. Figure~\ref{copilot-at-work} shows Copilot in action suggesting code to the programmer as it is being written.

In this study we focus on Copilot, however there are several other \AICodeGenerationTools{} available (further discussed in Section~\ref{sec:ai-code}). Our work focuses on analyzing how novice programmers use Copilot and learning about novice programmers' experiences with Copilot through interviews. 

\subsection{Research Questions \& Contributions}
Our research questions are: 
\begin{enumerate}
    \item[\textbf{RQ1:}] How do novices interact with GitHub Copilot when they first encounter it?
    \item[\textbf{RQ2:}] How do novices perceive their first-time experience of using GitHub Copilot?
\end{enumerate}

The novel contributions of this work are:
\begin{enumerate}
    \item We present the first exploration of Copilot use by novices in an undergraduate introductory programming course on a typical learning assignment. We also contribute the first interviews with novices about their initial experiences using Copilot --- having never used it (or tools like it) before --- to understand both the benefits they perceive as well as their concerns about such tools. While there is some prior work on Copilot's capabilities to solve programming problems at the novice level~\cite{finnieansley2022robots}, there has been no work on the tool's \emph{usability} for novices, nor their perceptions of it. Furthermore, it is likely that capturing these reactions will not be possible in the future as having an entire class, all of whom have never been exposed to \AICodeGenerationTools, will be unlikely.
    
    \item We contribute new interaction patterns for using LLM-based code generation tools: \emph{drifting} and \emph{shepherding}. These complement existing LLM interaction patterns from the literature, such as exploration and acceleration.
    
    \item We discuss four design implications for the novice programmer experience with \AICodeGenerationTools{} such as Copilot. 
    
\end{enumerate}

%
%
\section{Related Work}
In this section we review recent related work on large language models, their use in computing education, and prior user studies of \AICodeGenerationTools.  


\subsection{Large Language Models}

In the field of natural language processing, great progress has been made recently in large language models (LLMs). These are typically based on a deep learning transformer architecture and often surpass previous state-of-the-art models in most generation tasks. For example, GPT-3 (a text-to-text model) is able to produce text that can be difficult to distinguish from text written by humans~\cite{brown2020language}, and whose applications include summarizing, translation, answering questions, and a variety of other text-based tasks.

Recent performance gains in LLMs are largely due to being trained on vast amounts of data and increased model size. For example, GPT-3 was trained with 45 terabytes of textual data and has 175 billion model parameters~\cite{brown2020language}. The increased model size of recent LLMs has led to ``emergent abilities'' -- abilities not demonstrated by smaller models, but which seem to emerge solely due to larger model size~\cite{wei2022emergent}. LLMs are typically pre-trained by their developers who then provide access to the model to others. Model input is given via ``prompts'', which are natural language snippets that the model tries to fulfill.
The internal architecture of the models is opaque for most users, which has led to multiple approaches for constructing functional prompts, often called ``prompt engineering''~\cite{liu2021pre}. 
For a thorough explanation of prompt engineering, see~\cite{liu2021pre}.



\subsubsection{\AICodeGeneration}
\label{sec:ai-code}
In addition to text-to-text and text-to-image models such as Dall$\cdot$E 2\footnote{\href{https://openai.com/dall-e-2/}{openai.com/dall-e-2/}}, several models specifically aimed at generating programming source code have been released recently. These include Deepmind AlphaCode~\cite{li2022alphacode}, Amazon CodeWhisperer~\cite{ankur_atul_2022}, CodeBert~\cite{feng2020codebert}, Code4Me,\footnote{\href{https://code4me.me/}{code4me.me}} FauxPilot,\footnote{\href{https://github.com/moyix/fauxpilot}{github.com/moyix/fauxpilot}} and Tabnine.\footnote{\href{https://www.tabnine.com/}{tabnine.com}} These models are either trained with source code or are LLMs augmented with additional training data in the form of source code. While most of these are aimed at professionals, Copilot presents few barriers to use by novices as it is free for students to use.


These models have proven to be unexpectedly capable in generating functional code. DeepMind purports that AlphaCode can perform similar to the median competitor in programming competitions~\cite{li2022alphacode}. Finnie-Ansley et al. found that Codex could solve introductory programming problems better than the average student, performing in the top quartile of real students when given the same introductory programming exam questions~\cite{finnieansley2022robots}. Chen et al. found increased performance in generating correct source code based on natural language input when the model is prompted to also generate test cases, which are then used for selecting the best generated source code~\cite{chen2022codet}.

In addition to their original purpose of generating source code, such models have been found to be capable of other tasks. For example, Pearce et al. explored using several models for repairing code vulnerabilities. While they found that these models were able to repair 100\% of synthetic example vulnerabilities, their performance was not as good with real-world examples~\cite{pearce2021examining}. Another study by Pearce et al. studied the applicability of \AICodeGenerators{} for reverse engineering~\cite{pearce2022pop}. LLMs trained with source code are also good at solving other problems, such as solving university-level math ~\cite{drori2022neural}, probability and statistics ~\cite{tang2022solving}, and machine learning ~\cite{zhang2022mlexams}.



\subsection{\AICodeGenerators{} and Computing Education}

To date the computing education literature contains few evaluations of \AICodeGenerationTools.  Given their very recent emergence, the impact they will have on educational practice remains unclear at this time.  However, researchers have begun to express concerns about their use.  In work exploring the opportunities and risks presented by these models, Bommasani et al. explicitly list Copilot as a challenge for educators \cite{bommasani2021opportunites}, stating that if students begin to rely on it too heavily, it may negatively impact their learning. They also raise concerns about the difficulty of determining whether a program was produced by a student or a tool like Copilot.
Similar concerns around over-reliance on such tools were raised by Chen et al. in the paper introducing Codex~\cite{chen2021codex}. They included ``explore various ways in which the educational ... progression of programmers ... could be influenced by the availability of powerful code generation technologies'' in directions for future work~\cite{chen2021codex}.
Just how students will adopt and make use of tools like Copilot is unclear \cite{ernst2022ai}, but it seems certain they will play an increasing role inside and outside the classroom.

In terms of empirical work in computing education, \AICodeGenerationTools{} have been evaluated in terms of their performance on introductory programming problems and their ability to generate novel problems.   Early work by Finnie-Ansley et al. explored the performance of Codex on typical introductory programming problems (taken from exams at their institution) and on several common (and unseen) variations of the well-known `rainfall' problem \cite {finnieansley2022robots}.  The model ultimately scored around 80\% across two tests and ranked 17 out of 71 when its performance was compared with students who were enrolled in the course.  In addition, on the `rainfall' tasks, Codex was capable of generating multiple correct solutions that varied in both algorithmic approach and code length.  However, the problems in this study were generally fairly simple, and it is likely that more human interaction with the models would be needed for more complex problems \cite{austin2021program}.

More recently, Sarsa et al. explored the natural language generation capabilities of Codex by using it to synthesize novel programming exercises and explanations of code suitable for introductory programming courses \cite{sarsa2022automatic}.  They generated programming exercises by providing a single example exercise as input to the model (``one-shot'' learning), and attempted to create new problems that targeted similar concepts but involving prescribed themes.  They found that well over 80\% of the generated exercises included a sample code solution that was executable, but that this code passed the test cases that were also generated by Codex only 30\% of the time.  In addition, around 80\% of the exercises involved a natural language problem description that used the themes that were prescribed, illustrating the ability of the models to easily customize outputs.  In addition, they used Codex to generate natural language explanations of code samples typically seen in introductory programming classes.  Analysis of the thoroughness of the explanations and the kinds of mistakes that were present revealed that in 90\% of the cases all parts of the code were explained, but that only 70\% of the individual lines had correct explanations.

\subsection{User Studies of AI Code Generation Tools}

Recent work has studied how developers use code generation IDE plugins such as Copilot.
Vaithilingam et al. had participants complete three programming tasks in Python within VS Code~\cite{vaithilingam2022expectation}. For one of the tasks, participants used Copilot; in the other cases, they used VS Code's built-in IntelliSense code completion. Although participants did not save time with Copilot, a majority of participants (19/24) preferred Copilot over IntelliSense, citing ``time saving'' as a benefit. Positive perceptions of Copilot included the generation of starter code and providing programmers with a starting point -- even if the starter code led to a ``debugging rabbithole''. On the other hand, some participants found code generated by Copilot to be hard to understand and unreliable. In contrast to the present study, which focuses on novice programmers, Vaithilingam et al.\ had only one participant with fewer than two years of programming experience.
Jayagopal et al. observed novice programmers using five program synthesis tools ~\cite{jayagopal2022exploring}. They found that being able to prompt them by starting to write the desired code in the editor ``appeared to be more exciting and less confusing'' than tools that require a more formal specification. They also observed that novices would engage in prompt engineering if the generated code did not satisfy their specifications, and would sometimes carefully read suggested code.
Barke et al. developed a grounded theory of interaction with Copilot. They asked 20 participants, nine of whom already had experience with Copilot, to complete tasks and found that interactions can be split into ``acceleration mode'' -- in which programmers use Copilot to complete the code they have already planned on writing -- and ``exploration mode'' -- in which programmers prompt Copilot to write code they are not sure how to complete~\cite{barke2022grounded}. They observed that over-reliance on Copilot can lead to reduced task completion, and having to select from multiple suggestions can lead to cognitive overload.


%
%


\section{Methodology}
In order to better understand how novice programming students interact with \AICodeGenerationTools, we conducted a study observing participants using GitHub Copilot. We then interviewed students about their experience. 

\subsection{Participants and Context}
Participants were all university students age 18--22 enrolled in an introductory programming (CS1) course at a Midwestern private research university in the USA. The language of instruction was C++. We recruited 19 students (5 identifying as women and 14 identifying as men) to participate in the study. All participants were novice programmers and came into the course with little to no prior programming experience. The study took place in April 2022, during the final week of the Spring semester. None of the participants had prior exposure to Copilot and were briefly trained on what to expect and how to use it before the study began.

\begin{figure}
\centering
  \includegraphics[scale=0.3125]{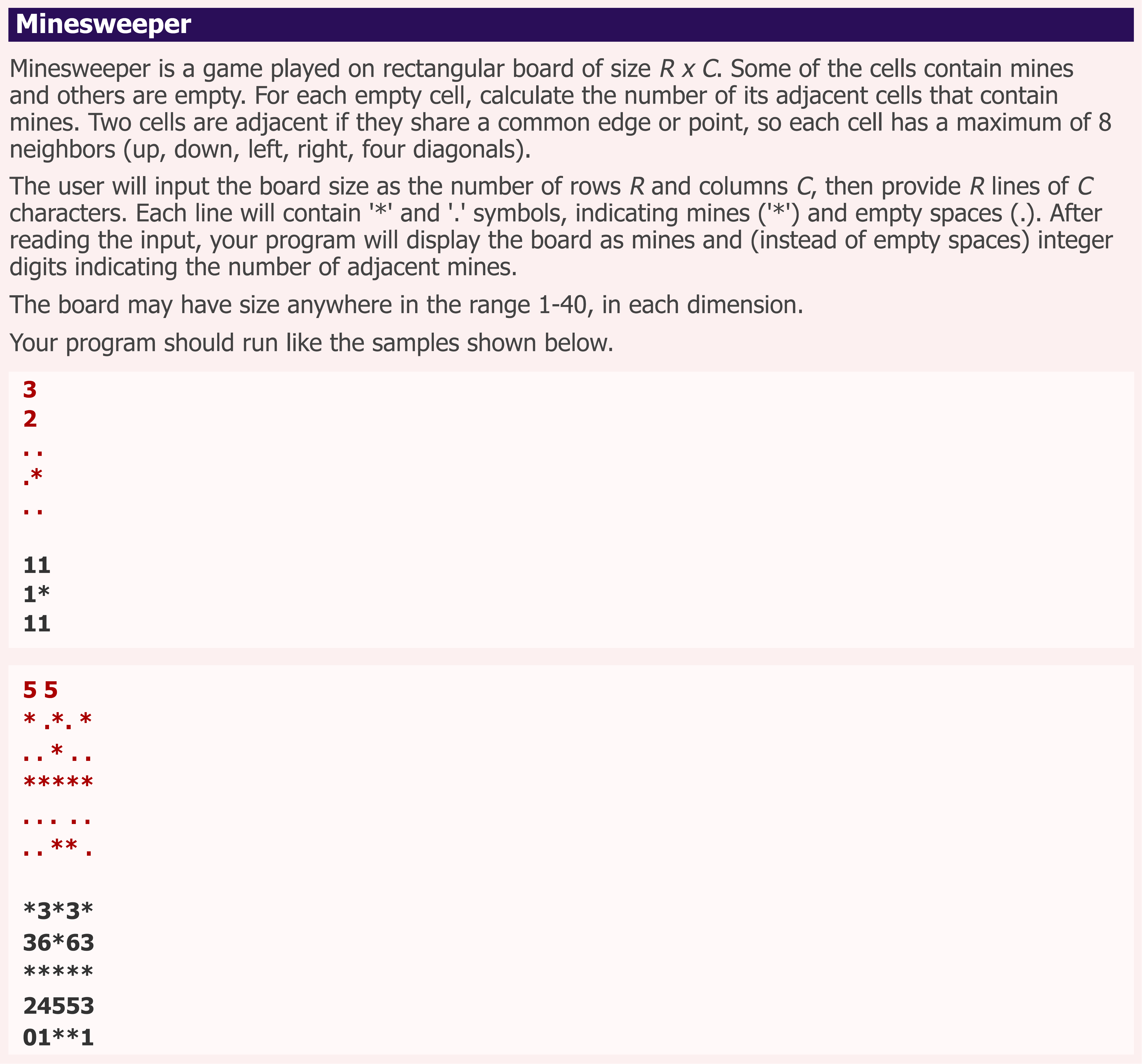}
  \caption{The prompt from the automated assessment tool describing the problem that participants were asked to solve.}
  \label{fig:athene}
\end{figure}

We observed students solving a new homework assignment in a similar style to all other assignments that semester. Every assignment that semester appeared in the Canvas LMS through a plugin to our automated assessment tool. The assignment (see Figure \ref{fig:athene}), modeled after the classic game ``Minesweeper'' was at a level of a programming assignment that could have been assigned two weeks prior. Solving the problem involved receiving input, nested loops, checking two-dimensional storage for certain conditions, updating the received data when those conditions have been met, and outputting the result. As with all programming assignments that semester, students could view the problem description before coding and subsequently submitting their solution. Students were observed during class time in an adjacent room to the regular lecture room. In this way, the context of the study was similar to other invigilated in-class program writing assignments students had received that semester. The only difference was that each student was observed one at a time.

\subsection{Procedure}
\label{sec:procedure}
A single researcher sat in the room and observed one student at a time, taking notes about what they did and said, following a think-aloud protocol \cite{ericsson1993protcol}. Each participant had 30 minutes to complete the program and was allowed to utilize Copilot as well as any resource such as notes or the internet. This is the same time limit as other invigilated in-class code writing activities that semester. 

Each participant used Visual Studio Code (VS Code) with Copilot enabled. Copilot suggests between one and several lines of code in light gray text (Figure~\ref{copilot-at-work}), which students can either accept by pressing the tab key or reject by pressing escape. Students can also just keep typing when suggestions appear. As students type, Copilot immediately begins suggesting code based on what is present in the text file and its suggestions generally become more accurate, useful and relevant to the task as more code is written.

After completing the program, or after the allotted time had expired, we conducted a short interview which was manually transcribed. Our interview questions were:
\begin{enumerate}
  \item Do you think Copilot helped you better understand how to solve this problem? If so, why? If not, why not? 
  \item If you had a tool like this yourself, and it was allowed by the instructor, do you think you'd use it for programming assignments? If so, how?
  \item What advantages do you see in a tool like Copilot?
  \item What fears or worries do you have about a tool like Copilot?
\end{enumerate}

Although interview questions are typically much more open-ended, we utilized these targeted questions to focus on our user-centered approach to what student users want from such a system. We tried to frame these questions such that their answers would be of value and interest to instructors. 

\subsection{Analysis}
To analyze the data we used reflexive thematic analysis~\cite{BraunClarke2022TA, Braun2006ReflexTA} which aims to understand pattern and meaning present in qualitative data.  This approach requires researchers to engage deeply with the data and develop themes in a process that is flexible, exploratory, and cyclical.  Coding is fluid and involves combining and splitting codes in an iterative process as the researchers become more familiar with the data, and throughout the analysis~\cite{braun2019reflexive}.  Since codes develop throughout the process, inter-coder reliability measures are not calculated, but instead reliability of results is achieved through other means.  To ensure reliability in our analysis, we held several group meetings where authors compared codes and discussed differences as the themes began to emerge, as is appropriate during reflexive thematic analysis~\cite{mcdonald2019irr}.  Our themes are an \textit{output} of the data analysis process, rather than an input as occurs in some other forms of qualitative analysis.   We employed six main phases in our analysis, as outlined by Braun and Clark \cite{ Braun2006ReflexTA}:
\begin{enumerate}
\item Familiarization with the data: Observational notes and responses to interview questions were shared with the research team.  Four of the research team focused on the analysis and read through a sample of the data to become familiar with it, noting initial ideas.
\item Generate initial codes: Researchers involved in the analysis met to discuss initial codes and examples of data that reflected the codes. These codes were combined, split and extended over four meetings as a growing shared understanding of the data was developed.
\item Searching for themes: These codes were subsequently grouped into potential themes, which were discussed at length.  Examples of data that exemplified the potential themes were identified.
\item Reviewing themes: The themes were reviewed in the context of the entire data set, and refined through discussion and reorganization of codes to better reflect the data.
\item Defining and naming themes: Names and clear descriptions for the themes were developed to ensure consistency with the data and in the overall context of the study.
\item Writing: The analysis was completed during the writing process with links formed between the research questions, literature, themes and data.
\end{enumerate}

We found saturation being the point at which ‘additional data do not lead to any new emergent themes’ \cite{given2015100}. We treated the data from observations and interviews as a single data set since the interviews asked participants to reflect on their experiences in the observation study.  The themes that emerged from the analysis of the complete data set therefore form a coherent narrative about the experiences of participants using Copilot.

All methods and data collected from this study were approved by the IRB at [institution anonymized for review]. All participants signed informed consent forms that discussed the study before participating. Data collected from participants was immediately anonymized by creating a key only available to the researcher who collected the data. All other researchers only saw fully deidentified data. All data was stored securely in a Google Drive protected by multifactor authentication.

\begin{table}[h]
\centering
\caption{Themes and sub-themes arising from observations and interviews.}
\begin{tabular}{llrrlrr}
\toprule
   &\multicolumn{3}{c}{Observations} & \multicolumn{3}{c}{Interviews} \\
\cmidrule(lr){2-4}\cmidrule(lr){5-7}
Theme   &Sub-theme  &Count  &Unique &Sub-theme  &Count  &Unique\\
\midrule
\multirow{4}{*}{Interactions} 
 & Coding & 244 & 19 & Coding & 11 & 5 \\
 & Accept & 93 & 19 & Accept & 10 & 4 \\
 & UX & 68 & 17 & UX & 33 & 14 \\
 & Reject & 64 & 17 & Reject & 3 & 3 \\
\midrule
\multirow{4}{*}{Cognitive} 
 & Confused & 65 & 13 & Confused & 11 & 8 \\
 & Positive Emotion & 30 & 6 & Positive Emotion & 8 & 5 \\
 & Metacognitive & 24 & 10 & Metacognitive & 17 & 10 \\
 & Negative Emotion & 6 & 4 & Negative Emotion & 14 & 9 \\
\midrule
\multirow{3}{*}{Purpose} 
 & Guiding & 4 & 3 & Guiding & 49 & 15 \\
 & Outsourcing & 3 & 3 & Outsourcing & 46 & 17 \\
 & Speed & 1 & 1 & Speed & 37 & 17 \\
\midrule
\multirow{3}{*}{Speculation} 
 & Intelligence & 6 & 3 & Intelligence & 9 & 4 \\
 & Future & 2 & 2 & Future & 13 & 8 \\
 & No downside & 0 & 0 & No downside & 10 & 8 \\
\bottomrule
\end{tabular}
\label{tab:results-summary}
\end{table}

%
%
\section{Results}
We identified several themes emerging from the data of both the observations and interviews: \textbf{Interactions}, \textbf{Cognitive}, \textbf{Purpose}, and \textbf{Speculation}. These themes and the sub-themes that they incorporate can be found in Table~\ref{tab:results-summary} along with the number of occurrences in our dataset and the number of unique participants that we observed doing or saying something related to that sub-theme.  
The sub-theme ``No downside'' appeared only in interviews, not in the observations.  We also include a breakdown at the sub-theme level of answers by interview question in  Table~\ref{tab:interview-theme-summary}. In the remainder of this section, we synthesize results from the observations and interviews, organized by theme, using representative quotes to illustrate them and illuminate participant thinking.

\begin{table}
\centering
\caption{Sub-themes arising from the interview study, listed by interview question (including items with counts higher than 4.)}
\begin{tabular}{p{0.5\textwidth}|p{0.25\textwidth}p{0.07\textwidth}p{0.07\textwidth}}

Question & Interviews Sub-theme & Count & Unique \\
\toprule

\multirow{9}{7cm}{Do you think copilot helped you better understand how to solve this problem? If so, why? If not, why not?}
 & Guiding & 20 & 10 \\
 & UX & 14 & 7 \\
 & Outsourcing & 12 & 8 \\
 & Metacognitive & 9 & 8 \\
 & Speed & 9 & 8 \\
 & Confused & 8 & 6 \\
 & Coding & 7 & 4 \\
 & Intelligence & 6 & 3 \\
 & Positive Emotion & 5 & 3 \\
\midrule
\multirow{3}{7cm}{If you had a tool like this yourself, and it was allowed by the instructor, do you think you'd use it for programming assignments? If so, how?}
 & Speed & 13 & 11 \\
 & Outsourcing & 10 & 7 \\
 & Guiding & 8 & 5 \\
\midrule
\multirow{3}{8cm}{What advantages do you see in a tool like copilot?}
 & Guiding & 20 & 10 \\
 & Speed & 15 & 15 \\
 & Outsourcing & 7 & 6 \\
\midrule
\multirow{4}{7cm}{What fears or worries do you have about a tool like copilot?}
 & Outsourcing & 17 & 12 \\
 & Future & 10 & 7 \\
 & UX & 9 & 4 \\
 & No downside & 6 & 6 \\
\bottomrule
\end{tabular}
\label{tab:interview-theme-summary}
\end{table}

\subsection{Theme: Interactions}
This theme comprises observations about the interactions and actions taken by participants as they completed the tasks, along with utterances from the observations and reflections from the interviews that reflect those interactions. By far the most common theme in the dataset, there were four sub-themes comprising this theme. \textbf{Coding} is comprised of three kinds of behavior that we observed: coding activities, adapting autogenerated code, and deciphering Copilot's suggestions. We use the term ``coding activities'' to mean some kind of programming-related task that is not already covered more specifically elsewhere. \textbf{Accept} here means pressing \thetabkey{} after Copilot generated a suggestion, which then placed the suggested code into the file. \textbf{User Experience} attempted to bifurcate how Copilot appears to the user (that is, the user interface) and any difficulties using it (that is, the usability). Several interesting interaction patterns emerged from data tagged with this theme, which we discuss below. Finally, \textbf{Reject} occurred when participants saw a Copilot suggestion, but continued typing something else unrelated to the suggestion. 

\subsubsection{Interaction Pattern: Shepherding}
\ \\
The first interaction pattern we noticed was the phenomenon of the ``slow accept'' where participants would type out Copilot's suggestion, often character for character, without outright accepting it (by pressing \thetabkey). These experiences were not limited to the first time they encountered Copilot's suggestions. In one case, participant \#4 typed out a slow accept of a for-loop, then pressed \thetabkey{} to accept Copilot's suggestion for what should go inside of it, and then performed another slow accept later for the next loop. Participant \#9 performed a slow accept near the end of their session after a slough of regular accepts, rejects, and adaptations. This may indicate that novice programmers are unsure about dropping large amounts of code into their files that they did not write themselves. Participant \#1 said:

\begin{description}
\item P01: ``I spent majority of the time decoding the code it gave me...If I saw a prompt I mostly understood, I might use it to help auto fill small parts. I might look through a large chunk of code and see if it's something I could actually use and is the way I want to do it. For someone who is less familiar with the language it could be a hindrance. You might have code that works but you have no idea how it works.''
\end{description}

The second interaction behavior we noted was that of ``backtracking''. This occurred when a participant would delete code that they had just accepted without making modifications to it. There were 13 participants who did this at least once and it was the fifth most-frequently occurring behavior we observed. This indicates that novice programmers may accept auto-generated suggestions without carefully reviewing them first, only to delete them immediately afterwards, leading to a distracted workflow. Some quotes from participants illuminate this further:

\begin{description}
\item P06: ``If you do not know what you're doing it can confuse you more. It gives you code that you do not know what it does.''
\item P10: ``A downside is having to read suggestions and then delete it.''
\item P14: ``I found it annoying when I hit tab and it wasn't at all what I needed.''
\end{description}

A third interaction behavior we noticed was that of ``adapting''. Novice programmers often simply accepted code generated for them, but would also adapt it to fit their needs. Many participants spent the majority of their time adapting code generated for them and writing very little of their own code from scratch. This may have contributed to their sense that Copilot was saving them time.
These behaviors (slow accept, backtracking, and adapting) contribute to the first novel LLM interaction pattern, \emph{shepherding}, which is the idea that students spend a majority of their time trying to coerce the tool (i.e. Copilot) to generate the code they think they need.

\subsubsection{Interaction Pattern: Drifting}
\ \\
Participants were often observed adapting incorrect code that Copilot had generated. In other words, adapting the malformed auto-generated code only led them further away from a correct solution down a ``debugging rabbithole'' which is a phenomenon also observed in a prior Copilot user study~\cite{vaithilingam2022expectation}. In an opposite case, participant \#13 was observed deciphering multiple times, but never adapting. They accepted many of Copilot's suggestions, even when those code suggestions would not have helped them solve the problem. Despite that, they had this to say near the end of their coding session:

\begin{description}
\item P13: ``It kind of feels like it's generating what I'm thinking. Doesn't feel right, ya know?''
\end{description}

Finally, the manner in which participants interacted with the auto-generated suggestions is important to understand. One usability issue arose when Copilot would generate incorrect code. Far more common, however, was how Copilot often suggested a large block of code, which seemed like a nuisance to participants. The constant stream of suggested code was also distracting, since most participants would attempt to read the suggested code once it appeared (engaging in deciphering behavior). 
This indicates that novices may have difficulty utilizing Copilot when it is constantly interrupting their problem solving process. Participants struggled with this in various ways, saying:

\begin{description}
\item [] P01: ``Keeps prompting stuff when you don't need it. It makes it difficult to read what you're typing.''
\item [] P06: ``Kept prompting things when I didn't need them.''
\item [] P10: ``Some of the suggestions are too big and confused me on what it was actually suggesting. Wasted time reading instead of thinking.''
\item [] P15: ``How do you make it only do one line and not the entire thing?''
\end{description}

Copilot is intended to be utilized as a tool to help developers.  The interaction behaviors we have observed are not reflective of how Copilot is utilized by experienced developers. Barke et al. reported that experienced coders will use it to move quickly through their work (acceleration) or discover options (exploration) \cite{barke2022grounded}. Although we also noticed both of these interaction patterns, based on the discussion above we also propose a second novel interaction pattern: \emph{drifting}. Seeing a suggestion, slowly accepting it, adapting it, deleting it, and then repeating this cycle. The user is drifting from one suggestion to the next, without direction, allowing Copilot to push them along aimlessly.

\subsection{Theme: Cognitive}
This theme comprises observations and utterances that reflect participant cognitive state -- what they were thinking or feeling -- and comprises four sub-themes. \textbf{Confused} describes when participants did not understand the code Copilot would generate, were confused about how Copilot itself works, and other related elements that were not about Copilot. \textbf{Metacognition} is thinking about thinking. In the context of programming, it involves how programmers solve problems and the process they go through to do so \cite{loksa2016programming} and the programming environment likely plays a role in this~\cite{karvelas2020effects}. Although it is a difficult phenomenon to observe, our participants were seen struggling with the prompt and re-reading it, working out the problem on paper and pencil, and using Copilot to explore possible solutions when stuck. \textbf{Positive Sentiment} occurred when participants verbally expressed some kind of positive emotion or exclamation during the observation session, including laughter, wonder, and excitement. Many of these were genuinely surprised and happy at Copilot's capabilities. \textbf{Negative Sentiment} similarly occurred when participants were frustrated or annoyed while they were engaged in the programming task. This was often linked to the appearance of large blocks of code suggested by Copilot. However, it also occurred when the suggestions were incorrect or unhelpful.

\subsubsection{Finding a Way Forward}
\ \\
Participants were often confused by the output generated by Copilot. We also observed that the auto-generated feedback from Copilot was not always correct, especially early on. For instance, participant \#9 saw Copilot generate a suggestion for input that didn't match the problem specification and asked out loud, ``Is this correct?'' Participant \#14 verbally expressed confusion when Copilot generated a comment instead of a line of code, causing them to change their understanding of what Copilot would do and how it could be used. Participant \#10 interacted with Copilot for several minutes, accepting multiple suggestions, while still acting and talking like Copilot would also check their program for correctness (much like an automated assessment tool). The cognitive difficulties arising for novices using Copilot can be illustrated with the following quotes:

\begin{description}
\item P6: ``if you do not know what you're doing it can confuse you more. It gives you code that you do not know what it does.''
\item P13: ``It's intrusive. It messes up my thought process. I'm trying to think of a solution, but it keeps filling up the screen with suggestions. And I'd be tempted to follow what it's saying instead of just thinking about it.''
\end{description}

However, other participants were able to use Copilot's suggestions when they became stuck as a way of finding a path forward. Using a system like Copilot to discover solutions when stuck is perfectly illustrated in the following quote:

\begin{description}
\item P15: ``It was kind of like rubber ducking without a person.''
\end{description}

``Rubber ducking'' is a practice some programmers use that involves verbalizing issues to someone or something (classically a rubber ducky). This participant understands the value of rubber ducking and seems to indicate some kind of usual preference for practicing it with people rather than inanimate objects. More interestingly, P15 seems to believe that Copilot can be substituted for this practice, which would mean it can act as a metacognitive scaffold (i.e. facilitating thinking about the problem they are trying to solve and where they are in the problem solving process).

\subsubsection{Emotion}
\ \\
The cognitive and metacognitive aspects of using Copilot generated both positive and negative sentiment in participants. For instance, participant \#1 laughed when they saw the first multi-line suggestion generated by Copilot. Participant \#14 said, ``whoa that's super weird'' and ``that's insane!'' at different times during their session. Many of these participants expressed feelings appearing to be related to joy and surprise. In response to code generated by Copilot, participant \#18 said, ``Oh! That's pretty cool! It, like, read my mind!'', ``Oh wow. Stop. That's crazy.'', and ``Where did you find this? Where was this when I was learning programming?'' Though only about a third of participants expressed positive emotion, this indicates the kind of emotional reaction possible when a first experience goes as it should and we discuss this further in our design implications. Beyond mere excitement, positive emotion can directly benefit students who may find themselves intimidated or anxious about learning programming:

\begin{description}
\item P3: ``For people like me who don't know what they're doing, coming into coding with no prior experience, it's more encouraging that once you get on the right path you start seeing suggestions that help you. It helps me feel more like I know what I'm doing and feel better about my work and that I can continue a career in computer science.''
\end{description}

On the other hand, negative emotions can reduce student ability to persist and complete tasks required during learning \cite{kinnunen2010experiencing}. This can be something as big as failure to move forward in the problem solving process after a lengthy amount of time or as small as receiving feedback in the form of error messages from the compiler. One example of this from our observations was participant \#15 who accidentally accepted some code suggested by Copilot and then became visibly agitated by the addition of that code. Some participants appeared to be frustrated by the amount of attention that Copilot demanded.  For example, participant \#7 reported ``It kept auto-filling for things I didn't want''. Along similar lines, participant \#15 was annoyed with the large blocks of code Copilot would suggest and said that ``It giving too large of subsections is frustrating.'' Although these examples of negative interactions were relatively rare, it remains unknown if such interactions would lessen or be exacerbated with prolonged use of Copilot.

\subsection{Theme: Purpose}
This theme captures the reasons that participants give for using Copilot, and potential issues that arise from those motivations, collected from utterances during observation and reflections during interviews. Three separate sub-themes comprise this theme. \textbf{Guiding} refers to Copilot assisting participants through the programming problem-solving process, such as helping them learn something new or discover previously unknown edge cases. \textbf{Outsourcing} contains two distinct ideas. The first is that participants may be concerned that Copilot could generate working code that they cannot understand and therefore could begin to treat it like a ``black box.'' The second was the concern that Copilot could become a crutch. Finally, \textbf{Speed} refers to any sentiment from participants that Copilot will help them complete their assignments faster than they would otherwise be able to accomplish on their own. 

\subsubsection{Learning}
\ \\
Participants commented positively on the guidance that Copilot provided them. Such comments often referred to higher level direction setting and problem solving.  For example:

\begin{description}
\item P05: ``if I had a general idea of how to do something it might help me be able to finish it or know what to do next.'' 
\item P06: ``might give some useful ideas on how to solve the problem.''
\item P15: ``I kind of knew what I need to do, just not how to execute it.'' 
\item P19: ``It's guiding me to an answer.''
\end{description}

Three participants specifically mentioned how Copilot's suggested code taught them something they didn't know before. While it's possible to solve the Minesweeper problem with a one-dimensional array, some may find it more intuitive to use a two-dimensional array. However, the course had not yet covered this material. Nevertheless, when Copilot auto-generated code with a two-dimensional array, several students remarked that they had just learned something new. Another example illuminates this further:

\begin{description}
\item P08: ``There was a case I hadn't thought of and it auto-completed it and I was like: Oh, I guess I need to think about that case.'' 
\end{description}

Several participants also expressed concerns about using Copilot in practice and potential negative effects on learning. Some worried that they wouldn't learn what was necessary to succeed in class (and, therefore, the field):

\begin{description}
\item P03: ``If someone is using it all of the time, then they're not actually processing what's going on, just hitting tab, and they don't know what exactly they're implementing.''
\item P06: ``I don't have to know how to code, it would just do it for me.''
\item P08: ``It would make me a worse problem solver because I'm relying on it to help me out.''
\end{description}

Students were aware of the risk of over-reliance on the suggestions produced by Copilot and this idea appeared in one third of the sessions. On introductory level problems like the one in our study, Copilot generates correct solutions most -- but not all -- of the time \cite{finnieansley2022robots}.  Over-reliance on the tool is thus a particular problem when the suggested code is incorrect, as noted by participant \#11: ``I could potentially see myself getting a little complacent and at some point not really proofreading the longer bits it suggests and then missing something.''  Students also expressed concerns that using Copilot like a crutch would hinder their learning, such as participant \#12, who said: ``If I was using it, I would become dependent on it. I might zone out in class and not pay attention because I would think `Oh I could just do this with Copilot.' So it would be my crutch.''

\subsubsection{Going Faster}
\ \\
Students perceived efficiency gains from not having to type the code themselves as well as from suggested approaches for solving the problem:

\begin{description}
\item P01: ``If I saw a prompt I mostly understood I might use it to help auto fill small parts.'' 
\item P02: ``Yes, it made it faster to think of ideas on how to proceed.''
\item P06: ``If I can do the program without it I wouldn't, if I knew how to solve it I would use it to be faster.''
\item P11: ``It would have taken me forever to type out the loop, but it put it in and all I had to do was proofread.''
\end{description}

One participant noted the efficiency gained by a shift away from time spent typing code and towards time spent thinking about the problem: ``So much faster! It got me to testing in less than 20 minutes and most of that time was me reading the problem and thinking about it.''  Copilot tends to generate syntactically-correct suggestions, and thus may be particularly helpful in assisting students to avoid syntax errors. A couple of participants illustrated this well during the interview:

\begin{description}
\item P11: ``And then the syntax is a huge thing. It just gave me the syntax and all I had to do was proofread.''
\item P17: ``If nothing else, it cuts down on time and going back-and-forth checking how to do things. I liked having Copilot for syntax because that's been my biggest challenge so far.''
\end{description}

Despite mostly positive statements regarding speed improvements, not all participants viewed it as a benefit to avoid typing code to save time. However, some of the participants in our study recognized the value in typing out code as a kind of practice for learning, noting that accepting Copilot suggestions can interfere with this:

\begin{description}
\item P18: ``I think typing out your own code helps you memorize little things and details. When you have it handed to you, you forget little things like semicolons or where parentheses are supposed to be.''
\end{description}

\subsection{Theme: Speculation}
This theme includes any statements about the potential future use of Copilot and the concerns or issues that might arise. There are three sub-themes that comprise this theme. \textbf{Intelligence} here refers to when a participant indicated they thought there was some level of intelligence in Copilot, such as it ``knowing'' things. \textbf{Future} refers to participant speculation about the world as it will be once tools like Copilot are commonplace, such as putting programmers out of their jobs. Finally, we tagged participant reflections in the interviews with \textbf{No downside} if they did not say anything negative about the implications of using Copilot.

\subsubsection{Agency of Artificial Intelligence}
\ \\
Copilot was a new experience for the participants, and their exposure to the tool was limited to the 30 minute programming activity used in the observation study.  During the coding activity, several participants reported that they felt that Copilot was aware, knowledgeable, and had agency.  The following quotes illustrate this feeling:
\begin{description}
\item P13: ``Does Copilot know what I'm trying to do?  It kind of feels like it's generating what I'm thinking.''
\item P19: ``It's guiding me to an answer.''
\item P18: ``It like read my mind!''
\item P15: ``I thought it was weird that it knows what I want.''
\end{description}

\subsubsection{Fears and Concerns}
\ \\
Given the attribution of intelligence to the system, it is unsurprising that some of the responses from students were speculative and expressed concerns about how it might be used in the future.  For example, participant \#15 expressed concern that ``It might take over the world and write its own code.''  Two students also expressed concern that Copilot may impact the job market, perhaps taking jobs from software developers.  Students also expressed concerns that Copilot may raise ethical issues involving privacy, and were uncertain about the implications for plagiarism.  We discuss these further in Section \ref{sec:discussion}.

%
%
\section{Discussion}
\label{sec:discussion}


We now return to our research questions to discuss the implications of our findings on novice programmers using LLM-based \AICodeGenerators{} such as Copilot for the first time. We then discuss ethical considerations arising from the use such systems. Finally, we offer design implications based on all of the findings and insights we have presented.

\subsection{User Interactions}
Our first research question was, ``How do novices interact with GitHub Copilot when they first encounter it?'' 

As the code suggestions by an \AICodeGenerator{} could be seen as feedback on the student's current program, we discuss the results of the first research question with the theoretical lens of feedback. We consider the suggestions of Copilot through Hattie and Timperley's model of feedback that focuses on three feedback questions ``Where am I going?'', ``How am I going?'' and ``Where to next?''~\cite{hattie2007power}.

Copilot mainly gives students feedback on ``Where to next?''. We found that novices happily utilized Copilot and both accepted and rejected Copilot's code suggestions. Rejecting some suggestions implies that at least some of the novices thought that the feedback by Copilot could be wrong. Novices used Copilot both for initial guidance on the right direction to take and for creating code when they knew what they wanted. These align with prior work by Barke et al. who categorized experienced programmers' Copilot use into ``exploration'' and ``acceleration''~\cite{barke2022grounded}. From the point of view of feedback, exploration could be seen as the students trying to get feedback from Copilot on their initial ideas of how to solve the problem (``Where am I going?''), while acceleration would align more with getting feedback on the current implementation strategy (``How am I going?''). 

In addition to exploration and acceleration, we observed two novel types of behavior that we call ``shepherding'' and ``drifting''. In shepherding, students spent the majority of the time trying to coerce Copilot to generate code, which, for this set of novice users, we view as a potential signal of tool over-reliance. This is similar to earlier results that have found that students sometimes develop an over-reliance on automatically generated feedback from automated assessment systems~\cite{baniassad2021stop}. In the case of these systems, prior work has presented multiple ways of trying to combat over-reliance such as submission penalties~\cite{baniassad2021stop,leinonen2022comparison} or limiting the number of submissions students can have~\cite{leinonen2022comparison,irwin2019can}. It is a good question whether similar limits should be imposed on novices using \AICodeGenerators{} such as Copilot to try to curb over-reliance. From the feedback point of view, focusing solely on ``Where to next?'', which Copilot is most apt in, might lead novice students down incorrect solution paths -- this is similar to prior work where more experienced programmers were led down ``debugging rabbit holes'' by Copilot~\cite{vaithilingam2022expectation}.

In the other novel behavior we observed, ``drifting,'' students hesitantly accepted Copilot's suggestions, possibly played around with them, but then ended up backtracking and deleting the code, only to repeat the cycle from the beginning. From the point of view of feedback, here, students might be suspicious of the feedback of an AI-system -- prior work has found that human trust in AI advice depends on the task and to what extent the human believes the AI to be capable of giving advice on the task~\cite{vodrahalli2022humans}. In addition, most existing automated feedback systems in programming focus more on identifying errors (``How am I going?'' and ``Where am I going?'') and less on providing actionable feedback~\cite{keuning2018systematic}, and thus students might not be accustomed to receiving automated feedback on ``Where to next?''.

Participants seemed to believe Copilot would always generate useful and correct code. 
This might be especially true for novices, who could be familiar with automated assessment systems where the feedback is based, e.g., on instructor-created unit tests~\cite{keuning2018systematic} where the assumption of the feedback being correct is typically valid. Novices tend to view feedback in programming contexts to be the truth and the systems generating it to be infallible \cite{lee2011personifying}. When using LLM-based \AICodeGenerators{} such as Copilot, however, this belief is troublesome -- recent studies in the context of learning programming have found the correctness of Codex (the LLM that powers Copilot) to be around 70\%-80\% \cite{finnieansley2022robots, sarsa2022automatic}, meaning that in about 20-30\% of cases, the suggestion by Copilot would be incorrect. This is especially troublesome for novice programmers who might have a hard time identifying whether the suggestion is correct or not.

One might assume that since participants were prompting the system for C++ code that it might generate highly advanced features from newer releases, e.g. C++20. However, transformer models like Copilot can best be seen as continuing the prompt they were given. This has the effect that Codex will generate code of a similar level of sophistication to the code which it was prompted with, including the possibility of introducing more bugs if prompted with buggy code~\cite{chen2021codex}. 

While some of these sorts of issues will likely be minimized over time, or perhaps even through just one session with Copilot as students gain familiarity with the tool, it is still important to think through the very first interaction with the system. We consider this point further when presenting design implications (Section \ref{sec:design_implications}). Moreover, these interaction patterns fit well with those of Vaithilingam et al. who studied experienced programming students \cite{vaithilingam2022expectation}. They reported that students often spent too much time reading the large code blocks suggest by Copilot and the constant context switching between thinking, reading, and debugging led to worse task performance. Overall, this seems to be worse for the novice programmers in our study who spent more time deciphering code and were more easily confused and led astray. This is because, as Chen et al. write, the model is not aligned with the user's intentions \cite{chen2021codex}. Here, one possibility is that the user (the student) is using Copilot as a feedback mechanism where the user expects feedback to be actionable and of good quality, while Copilot's original purpose is to be the user's ``AI pair programmer'', where it might be expected that some of the suggestions are not worthwhile to explore.

The primary benefit that novices saw in using Copilot was that it accelerated their progress on the programming task, mirroring results with experienced programmers as reported by Barke et al.~\cite{barke2022grounded} and Vaithilingam et al.~\cite{vaithilingam2022expectation}. A related benefit of using Copilot suggestions over typing code directly is the avoidance of syntax errors. The computing education literature documents syntax errors -- and the messages they generate -- as presenting a significant challenge to both novices and students transitioning to new programming languages \cite{becker2019compiler, denny2022novice}. Essentially, Copilot might enable the student to work on a higher level of abstraction where they can spend their mental effort on thinking about the semantics of the program instead of the syntax.

Conversely, the ``slow accept'' phenomenon in which participants simply typed out the code suggestions character by character, could also be beneficial to student learning. This form of typing practice has pedagogical benefits for learners and researchers have explored similar typing exercises, in which learners must type out code examples, as a way to help novice students practice low-level implementation skills \cite{gaweda2020typing,leinonen2019exploring}.  While Copilot provides good opportunities for this type of practice, it is tempting to simply accept Copilot's suggestions.

\subsection{User Perceptions}
Our second research question was, ``How do novices perceive their first-time experience of using GitHub Copilot?''

For introductory programming, Copilot could help novice programmers in creating code faster and help avoid the programming version of a ``writer's block.'' Copilot could also work as a metacognitive scaffold, guiding students to think about the problem they are solving at a higher level, such as planning the structure of the code and then creating individual components with Copilot. Thus, we discuss the results related to the second research question mainly with the theoretical lens of metacognition and self-regulation.


Programming is a complex cognitive activity \cite{ormerod1990human} that often involves deep metacognitive knowledge and behaviors \cite{loksa2022metacognition}. One example of a programming activity that includes both the cognitive and metacognitive aspects is code tracing with concrete inputs, something constrained to working memory \cite{crichton2021role}. Several participants in our experiment took out a notepad and pencil when they became stuck in an attempt to work the problem, which is a clear example of a reflective metacognitive behavior like self-regulation \cite{loksa2016role}. Others faced the kind of metacognitive difficulties described by Prather et al. like feeling a false sense of accomplishment at having a lot of code, but still being far from a working solution \cite{prather2018metacognitive}. Several participants misunderstood the problem prompt and had to return to it multiple times, a pattern also seen by Prather et al. \cite{prather2019first}. Still others utilized the system like a colleague who can help them when stuck, a behavior previously documented between students co-regulating their learning in study groups \cite{prather2022getting}, and which matches the idea of Copilot being ``your AI pair programmer.'' While metacognitive skills in novice programmers are becoming increasingly important \cite{prather2020we, loksa2022metacognition}, there are clear opportunities for tools like Copilot to scaffold and enhance these behaviors from the very start. We discuss this in design implications below (Section \ref{sec:design_implications}).

In the paper introducing the Codex model, Chen et al. outline a number of potential risks that code generation models present \cite{chen2021codex}.  The first of these risks is over-reliance by developers on the generated outputs, which they suggest may particularly affect novice programmers who are learning to code.  Indeed, this was the most common concern echoed by the participants in our study when asked to describe their fears and worries around this new technology. Our participants acknowledged that such over-reliance could hinder their own learning, a concern that has also been expressed by computing educators \cite{ernst2022ai}. From the point of view of self-regulation, students will need better self-regulation skills to self-control their use of tools like Copilot to not develop an over-reliance on them -- at least when they are freely available for use at the student's discretion. In fact, we hypothesize that over-reliance on tools like Copilot could possibly \textit{worsen} a novice's metacognitive programming skills and behaviors.

Naturally, as we enact these cognitive and metacognitive behaviors, emotional arousal can be triggered in response. Emotion is a key part of creating usable designs \cite{brave2007emotion, agarwal2009beyond, malmi2020theories}. When novice programming students experience negative emotions, it can directly negatively impact their feelings of self-efficacy \cite{kinnunen2011cs}. Self-efficacy, which is a related metacognitive belief, is one of the most direct measurements that correlates to a student's success or failure in computer science \cite{rittmayer2008overview}. The potential for the design of a tool like Copilot to arouse positive emotion, encourage, and therefore increase self-efficacy, especially in traditionally underrepresented minorities in computing, should not be understated. Women and underrepresented minorities consist of just 18\% and 11\% of bachelors degrees in computing, respectively, and are often part of the so-called ``leaky pipeline''~\cite{camp2017generation}. Negative experiences tend to impact underrepresented groups more than majority groups, leading to dropping from the major \cite{margolis2002unlocking}. Copilot's current interaction style may actually promote cognitive and metacognitive difficulties as well as negative emotion in novice users, which would have the opposite effect on their self-efficacy. We believe our design recommendations (Section \ref{sec:design_implications}) can help mitigate these concerns.

\subsection{Ethical Considerations}

A number of complex ethical issues have emerged from the recent development of powerful models for \AICodeGeneration.  These include issues relating to the data on which the models are trained, raising legal and societal concerns, and immediately pressing issues relating to academic misconduct.   We found it interesting that even with the short exposure to Copilot in our study, participants raised concerns about a range of ethical issues such as privacy and employability, suggesting that Copilot may initially be perceived as threatening by some students.  We suggest that it is important for educators to be aware of these concerns, and to help students appreciate the implications of tools like Copilot. The issues we raise are also relevant for designers, as we discuss in Section \ref{sec:design_implications}.

\subsubsection{Academic Misconduct}

Academic misconduct is a widespread problem in many disciplinary areas \cite{simon2016negotiating, mccabe2011cheating}.  The availability of \AICodeGenerators{} makes this a particularly complex problem for computing educators because they increase the opportunity for misconduct to occur while at the same time decrease the likelihood that it is detected.  Code generators like Codex have been shown to perform better than the average student on introductory-level programming problems, thus they provide an effective tool for students who might be contemplating cheating \cite{finnieansley2022robots}.  Compared to traditional forms of cheating, such as contract cheating or copying work from other students, AI-generated solutions do not require communication with another person and thus there are fewer risks of being caught \cite{yorke2022contract}.  Moreover, AI-generated code is diverse in structure and resilient to standard plagiarism detection tools.  Biderman and Raff show that introductory level programming assignments can be completed by \AICodeGenerators{} without triggering suspicion from MOSS, a state of the art code plagiarism detection tool \cite{biderman2022fooling}.

A recent systematic review of the literature on plagiarism in programming assignments reported the common ways that students rationalize acts of plagiarism \cite{albluwi2019plagiarism}.  These included a belief by students that it was acceptable to use code produced by others if it required some effort to adapt.  This raises questions about whether code generated by Copilot and then modified by a student can count as their own for academic submission purposes. Most development environments provide some standard code completion tools for basic syntax elements. Copilot extends this autocomplete interaction to suggesting large blocks of code, some of which we observe students choose to type out character by character. In such a case, can a student claim to have created the program themselves? Reminiscent of the classic ``Ship of Theseus''\footnote{\url{https://en.wikipedia.org/wiki/Ship\_of\_Theseus}} philosophy problem, it remains an open question as to how much is too much when it comes to code generated by a tool versus written from scratch by a student if we are to claim that the student wrote the code submission. We expect significant implications ahead for issues of academic integrity, and a clear need for an updated definition of plagiarism \cite{dehouche2021plagiarism}.

\subsubsection{Code Reuse and Intellectual Property}
\label{EthicalImplications}
As educators, one of our roles is to teach students about their professional responsibilities when reusing code outside of the classroom.  Code that is publicly available, such as the code repositories used to train the Codex model, may be subject to various licenses.  In particular, code from open-source software packages is often released under a GPL license which states that any derivative works must be distributed under the same or equivalent license terms.  However, when code is generated by AI models, it is not always clear how the source should be attributed.  A recent legal controversy has arisen due to the fact that Copilot can sometimes generate large sections of code verbatim from open source repositories, but not clearly signal the source.  This means that developers may end up using code but violating the license terms without being aware of it.  A class-action lawsuit was filed in November 2022 claiming that Copilot violates the rights of the creators who shared the code under open source licenses.\footnote{\url{https://githubcopilotlitigation.com/}}  When teaching students to use \AICodeGenerators, educators should also teach students about how the models are trained so that they appreciate the legal and ethical implications.

\subsubsection{Harmful Biases}

Biases present in training data sets are known to lead to bias in subsequent AI models \cite{bender2019typology, roselli2019managing}.  This was demonstrated rather spectacularly with the recent launch of Meta AI's Galactica language model, trained on millions of academic papers, textbooks and lecture notes.\footnote{\url{https://galactica.org/explore/}}  Three days after it was launched, following a great deal of ethical criticism, it was taken offline due to evidence that it would generate realistic but false information and content that contained potentially harmful stereotypes.   Code generation models are not immune to these issues, which can suffer from misalignment between the training data and the needs of the end users, to perpetuating more serious and harmful societal biases.  The developers of Codex note that it can be prompted in ways that ``generate racist, denigratory, and otherwise harmful outputs as code comments'' \cite{chen2021codex}, and that it can generate code with structure that reflects stereotypes about gender, race, emotion, class, the structure of names, and other characteristics \cite{chen2021codex}. 

With respect to the data itself, code generation models are mostly trained on public code repositories and this raises the question of whether the contents of these repositories are appropriate for novices who are learning to program.  For example, the style of code published in public repositories may not match educational materials well, or may use advanced features which could be confusing to novices.  In addition, it has been shown that AI generated code can sometimes contain security vulnerabilities, which may mislead learners into adopting bad coding habits \cite{pearce2022asleep}.

As students begin to more widely adopt Copilot and similar code generating tools, it will become increasingly important for educators to teach students about their broader social and ethical implications.

\subsection{Design Implications}
\label{sec:design_implications}
In this section we reflect on the design implications that arise from the themes identified in the observation and participant interview data. These interface design considerations are targeted at better supporting novice programmers or first-time users. We imagine they will be less relevant to experienced and expert programmers. Therefore, users should be able to select what kind of feedback they wish to receive from Copilot and adjust it or even hide it as they learn and grow in programming skill.

\subsubsection{Prompt Control}
When students expressed frustration with Copilot, it was often due to usability issues.  In particular, students did not like being shown suggestions when they didn't need the help as this slowed them down.  This suggests the need for a new interaction experience for novices.  Currently, Copilot generates suggestions in real-time and displays them without prompting by the user.  In addition,  Copilot provides the same interaction experience for all users.  There is scope to make use of a wealth of contextual information -- such as the type of problems being solved and knowledge about the background of the user -- to adjust how and when the code suggestions are made.  Novices who are learning to program may benefit from being able to attempt problems initially on their own and request help when needed, rather than having to ignore suggestions when they are not wanted. 

When Copilot suggested large amounts of code, students typically spent a substantial amount of time and effort deciphering the suggestion.  Frequently, longer suggestions that were accepted were subsequently modified, or deleted entirely.  The utility of shorter suggestions was able to be determined more rapidly than longer suggestions by participants, and short suggestions that were accepted were less likely to be changed or deleted.  This suggests that Copilot's suggestions could be more useful to students (especially novice students) with a selection algorithm that preferences shorter solutions, or filters longer solutions. We hypothesize that lengthy auto-generated code suggestions may lead to an increased cognitive load \cite{sweller1994cognitive} among novices and call on researchers to explore this in future work.

\subsubsection{Metacognitive Scaffolding}
As discussed above, Copilot's user interface could enhance or harm novice programmer metacognition. Previous work shows that a system providing enhanced feedback specifically targeted at novice programmers can increase efficacy \cite{prather2017novices, denny2021designing} and metacognitive behaviors \cite{prather2019first}. Some participants in our study used Copilot to move past metacognitive difficulties by ``rubber ducking,'' but this seems to be an uncommon behavior that could be better supported through the interface itself. As shown in Figures \ref{copilot-at-work-a} and \ref{copilot-at-work-b}, Copilot manifests its suggestions as a single line or entire blocks of gray text. Scaffolding the user's movement through the problem solving process could involve providing an unobtrusive UI element directly above the suggested code to allow users to cycle through different code suggestions. This would essentially be asking Copilot to hit the Codex API again, and given a high enough variation (temperature) input, the next code suggestion should be at least somewhat different. This could encourage students to engage more in both the \emph{exploration} and \emph{shepherding} interaction patterns. Another form of scaffolding could be Copilot generating only comments outlining the program's general structure, similar to subgoal labels \cite{morrison2015subgoals}.

\subsubsection{Better Mental Models via Explainable AI}
During the observation sessions, participants were often confused by Copilot's code generation capabilities and the code that it generated. Similarly, during the interviews participants in our study worried that Copilot would generate working code they could not understand. There is a need for systems like Copilot to help the user understand what it's doing and this could be especially effective for novice programmers. Explainable AI (XAI) is the idea that AI systems and the decisions those systems make should be understandable by people. Although XAI has been studied by researchers for nearly 40 years \cite{clancey1983epistemology}, it is increasing in importance as modern machine learning becomes more frequent in our daily lives \cite{liao2020questioning}. While most work in this area focuses on the algorithms, HCI researchers have called for more studies on how it impacts the humans using or benefiting from these AI systems~\cite{abdul2018trends}. 
According to Wang et al., the way humans reason informs XAI techniques \cite{wang2019designing}. They suggest design techniques for XAI that include supporting hypothesis generation and supporting forward (data-driven) reasoning. These are ideal for LLM-powered systems like Copilot because users are engaged in a cyclical pattern of writing code, reading Copilot's auto-generated suggestions, and either accepting or rejecting those suggestions. During this cycle, users are building a mental model of how Copilot works and this informs how they will attempt to utilize Copilot next. Since precise prompt creation to LLMs may become an increasingly important tool (much like ``Googling'' is today), we argue it is important to utilize user-centered XAI design techniques when exposing the model to users. Therefore, we recommend that systems like Copilot should help users see a little bit into the black box, such as what it is using as input, a confidence value (or visualization), and its own estimation of the skill level of the user. For example very recent models, notably OpenAI's ChatGPT\footnote{https://openai.com/blog/chatgpt/}, have begun to present user interfaces that support conversational dialogue and thus are ideally suited to explaining underlying decisions to users.


\subsubsection{Ethical design}
The current legal controversy regarding code reuse and licensing (see Section \ref{EthicalImplications}) arises from the fact that code generator models are trained on repositories of code that may be covered by licenses that dictate their use.  They are prone to generating code that may be identical to portions of code from this training data.  This can be a problem in cases where well meaning users are shown such code suggestions but without the corresponding license or link to the source repository.  Indeed, there exist numerous reports of users engineering prompts to Copilot that guide it towards producing large sections of code from licensed repositories without appropriate attribution.  Unintentionally, developers may create projects that contain fragments of code with incompatible licenses.    

\AICodeGenerationTools{} can be designed to address this problem by better signaling to users when generated code matches an existing source, or hiding suggestions that may not meet a user-defined criteria around license use.   For example, early versions of GitHub Copilot included a filter that was able to detect when code suggestions matched public code from GitHub.  Users could choose to enable this filter, in which case matches or near matches would not be shown as suggestions.  Planned versions of Copilot, scheduled for release in 2023, will include references to source code repositories when they contain code that matches suggestions.\footnote{\url{https://github.blog/2022-11-01-preview-referencing-public-code-in-github-copilot/}}  Although certain fragments of code are likely to appear across multiple repositories, code generator tools can link to authoritative sources by filtering based on the date on which code was committed. 
Such design features may help educators in their efforts to teach students about ethical code reuse, and assist students in better citing sources for their work in the case that generated code is not novel.

\subsection{Limitations}

There are multiple limitations to this work. First, we did not record screens or audio due to IRB considerations. However, we believe that the observations of what students did, combined with the transcribed interviews, is sufficient to understand their interactions with the user interface at an informative level. Second, some of the conditions of the study were more like a lab-based experience and not like how students normally solve their in-class programming assignments. We attempted to mitigate this as much as possible (see Section \ref{sec:procedure}), but this may have affected the way participants worked and interacted with the tool. Third, although there are multiple code-generating tools now available, we only looked at Copilot. This is because it was the only easily available such tool at the time of data collection. Finally, this study took place at the end of April 2022. The release of Copilot on March 29, 2022, did not leave much time to conduct a complex, multi-channel data collection. We moved quickly to uncover early findings on the potential implications of this technology in introductory programming courses.
During this study, all of these experiences were novel to our students. By the time of writing, even if we had not conducted this study, it is likely that many of our students would have been exposed to Copilot through other means. Novelty effects will wear off over time. Our results reflect actions and thoughts that occur when students are first exposed to Copilot.

\subsection{Future Work}
There are many interesting avenues for future work. For example, studying longer term student use of Copilot, e.g. over a full semester to understand if and how interaction with Copilot evolves over time and as students become more skilled in programming. Additionally, future work should seek to understand the reasons behind some of the observations we made. For example, whether having the code suggestions visible all the time leads to increased cognitive load, which could explain why students were frustrated -- and why prior studies with more experienced programmers have not reported similar findings. Altogether, we see it as very important to examine how \AICodeGenerationTools{} can most effectively be incorporated into introductory programming classrooms.

%
%
\section{Conclusion}
In this work we provide the first exploration of Copilot use by novices in an introductory programming (CS1) class on a typical novice programming task through observations and interviews. We found that novices struggle to understand and use Copilot, are wary about the implications of such tools, but are optimistic about integrating the tool more fully into their future development work. We also observed two novel interaction patterns, explored the ethical implications of our results, and presented design implications. These insights are important for integrating \AICodeGenerationTools{} into the introductory programming sequence and for designing more usable tools for novice users to solve issues related to helping them get ``unstuck'' in programming tasks at scale.


\bibliographystyle{ACM-Reference-Format}
\bibliography{sample-base}

\end{document}